\documentclass[]{aa}
\usepackage{graphicx}
\usepackage{txfonts}
\usepackage{longtable, lscape}
\usepackage{natbib}

\begin{document}
\title{The dark halo of the Hydra I galaxy cluster: core, cusp, cosmological?  \thanks{Based
    on observations taken at the European Southern Observatory, Cerro Paranal, Chile,
    under the programs 082.B-0680, 076.B-0154, 065.N-0166, 065.N-0459.
    }}

\subtitle{Dynamics of NGC 3311 and its globular
cluster system}

\author{
T. Richtler    \inst{1}\and
R. Salinas       \inst{1,2}\and
I. Misgeld      \inst{3}\and
M. Hilker     \inst{4}\and
G. K.T. Hau              \inst{2}\and
A.J. Romanowsky \inst{5}\and
Y. Schuberth     \inst{6}\and
M. Spolaor        \inst{7}
}

\offprints{T. Richtler}

\institute{
Departamento de Astronom\'{\i}a,
Universidad de Concepci\'on,
Concepci\'on, Chile;
[tom,rsalinas]@astro-udec.cl
\and
European Southern Observatory,
Alonso de C\'ordova 3107, Santiago, Chile
\and
Sternwarte der Universit\"at  M\"unchen, Scheinerstr.1, D-81679,  M\"unchen, Germany;
imisgeld@mpe.mpg.de
\and
European Southern Observatory,
Karl-Schwarzschild-Str.2, Garching, Germany;
mhilker@eso.org
\and
UCO/Lick Observatory, University of California,
Santa Cruz, CA 95064, USA;
romanow@ucolick.org
\and
Argelander Institut f\"ur Astronomie,
Auf dem H\"ugel 71, 53121 Bonn, Germany;
yschuber@astro.uni-bonn.de
\and
Australian Astronomical Observatory, PO Box 296, Epping,
NSW 1710, Australia;
max.spolaor@gmail.com
}

\date{Received  / Accepted }

\abstract
{Some galaxy clusters  exhibit  shallow or even cored dark matter density profiles in their central regions rather than the predicted  steep or cuspy  profiles, conflicting
with  the standard understanding of dark matter. 
NGC 3311 is the central cD galaxy of the Hydra I cluster (Abell 1060).}
{  We  use  globular clusters around NGC 3311,
combined with kinematical data of the galaxy itself, to  
investigate the dark matter distribution in the central region
of Hydra I .}
{Radial velocities of 118 bright  globular clusters, based on VLT/VIMOS mask spectroscopy, are used
to calculate velocity dispersions which are well defined out to 100 kpc. NGC 3311 is the most distant galaxy  for which this kind of study has been performed.
We also determine velocity dispersions  of the stellar component from long slit spectroscopy of NGC 3311 with VLT/FORS1 out to 20 kpc. Moreover, we  present a new photometric model for NGC 3311,
based on deep VLT/FORS1 images in the V-band.
We 
search for a dark halo which in the context  of a spherical Jeans model  can reproduce the kinematical data. We also compare the radial velocity distributions
of globular clusters and planetary nebulae.}
{The projected stellar velocity dispersion rises from a central low value of about 185 km/s to 350 km/s at a radius
of 20 kpc.
The globular
cluster dispersion rises as well from 500 km/s at 10 kpc to about 800 km/s at 100 kpc, comparable to
the velocity dispersion of the cluster galaxies. A dark matter halo  with a core (Burkert.halo) reproduces
well the velocity dispersions of stars and globular clusters  simultaneously under isotropy. The central stellar 
velocity dispersions predicted by cosmological NFW halos are less good representations, while the globular
clusters allow a wide range of halo parameters. A suspected radial anisotropy of the stellar population as found in merger simulations aggravates the deviations. A slight tangential anisotropy permits better representations. 
 However, we find discrepancies with previous kinematical data, which
we cannot resolve and which may indicate a more complicated velocity pattern.   }  
{Although one cannot conclusively demonstrate that  the dark matter halo of NGC 3311 has a core rather than a cusp,
 a core seems to be  preferred by the present data. A more complete velocity field and an analysis of
the anisotropy is required to reach firm conclusions. }

\titlerunning{Dynamics of NGC 3311}
\maketitle
\section{Introduction}
Cosmological simulations of  dark matter halos of galaxies predict the
density profiles in the very inner regions to be {\it cuspy}, i.e. the density
reaches very high values for small radii, and the logarithmic
density slope takes the value $ -1$ (the NFW profile; e.g. \citealt{navarro04}) or even smaller, e.g.
\citet{bullock01}, \citet{diemand05}.
 Confronted with observations of
low-surface brightness galaxies,
this prediction has not been confirmed  (see \citealt{deblok10} for a review).
It has instead been found that the density profiles are better described
by having a {\it core}, i.e. the central logarithmic density slope is zero.
Brighter galaxies are baryon dominated in their centers, and therefore the
inner dark
matter density slopes of bright spirals and ellipticals are more difficult to determine. However, there is some evidence for cuspy halos in early-type galaxies \citep{tortora10}.\\
Since dark matter halos are self-similar,
one also expects cuspy dark halos on the larger scales of galaxy
groups and galaxy clusters. The best test objects  are
central cD galaxies with  central low surface brightness. For example,
 \citet{kelson02} investigated NGC 6166, the central galaxy of Abell 2199
by long-slit spectroscopy reaching out to 60 kpc. They found that the
observations were best represented by a halo with a large core rather
than being cuspy.
Moreover,  \citet{sand04,sand08}, using strong and weak lensing
models in combination with the kinematics of the central galaxies, found for
the clusters Abell 383 and MS 2137-23 shallow dark matter profiles which
are incompatible with the steep NFW slope. The same result emerged from a
study of the lens properties of Abell 611 \citep{newman09}. See also \citet{newman11} for
an improved analysis of Abell 383.

 These are  important observations, pointing towards some fundamental deficiency
of the present understanding of dark matter. It is therefore of high interest
to investigate this issue by using a variety of dynamical tracers.

In this context, NGC 3311, the central galaxy of Abell 1060, is one of the
most attractive targets. It is the nearest cD galaxy, has a low surface
brightness, and exhibits an extremely rich globular cluster system, (e.g. \citealt{wehner08}),
providing a wealth of very bright globular clusters which can be used as dynamical
tracers. 
 NGC 3311 is  probably the most distant galaxy where
 this kind of study can currently be performed.
 Here we measured radial velocities of about 120 globular clusters.
   Simultaneously, we use long-slit spectra of NGC 3311 itself
to investigate the kinematics of the stellar body. Our objective is to  
describe the stellar kinematics and the GC kinematics consistently with the same
dark halo model.
 
We adopt a distance  modulus  of 33.4 (a handy value close to the mean given by NED), corresponding to 47.9 Mpc and a scale of 232 pc/arcsec.

\section{Observations and data}
Observations and reductions of all data used here are extensively presented in parallel or
previous contributions. Here we give a few basic remarks only.
The globular cluster data have been taken with  VLT/VIMOS on Cerro Paranal in service mode.
We merged two  VIMOS programmes, using multi-slit masks with the medium resolution grism.
Details of the observations, data reduction, and derivation of radial velocities  are described
by  Misgeld et al. 2011 (submitted to A\&A).
  The VLT/FORS1 images in the V-band which provide the photometric model of NGC 3311,
are discussed in \citet{misgeld08}. 
 The long-slit spectra  have been obtained using VLT/FORS1
at  Cerro Paranal, using the grism 600B.
Observations and data reduction are  described in detail by Spolaor et al. 2010 (submitted to MNRAS).
A preliminary account of the long-slit data has been given by \citet{hau04}.

\section{Ingredients for a Jeans-analysis: Photometric model and kinematics}

\subsection{Photometric model for NGC 3311}

 We derive the light profile of
NGC 3311 by applying IRAF/ELLIPSE in an iterative manner, subtracting in each
iteration the contributing light of the neighbouring galaxy NGC 3309
until its residual light   is not longer noticeable.

Since we want to work with an analytical expression for the projected light which
also permits an analytical deprojection and  an analytical 
cumulative luminosity, we prefer a double beta-model for the
surface brightness (see also
\citealt{schuberth10}) for which we obtain:

\begin{equation}
\label{eq:light1}
\mu_V(R)=-2.5\log \left(a_1 \left(1+\left( \frac{R}{r_1} \right)^2 \right)^{\alpha_1} + a_2 \left (1+\left(\frac{R}{r_2}\right)^2 \right)^{\alpha_2} \right)
\end{equation}
with $a_1$\,=\,1.838$\times$10$^{-8}$, $a_2$\,=\,2.67$\times$10$^{-9}$, $r_1$\,=\,5\arcsec,
$r_2$\,=\,50\arcsec, $\alpha_1 = \alpha_2$\,=\,$-1.0$.  
To transform
 this surface brightness into  $L_\odot/pc^2$, one has to apply a factor $3.7 \times  10^{10}$ to the argument of the logarithm. An additional factor 1.27 corrects for the extinction in the V-band (\citealt{schlegel98}).

\begin{figure}
\begin{center}
\includegraphics[width=0.5\textwidth]{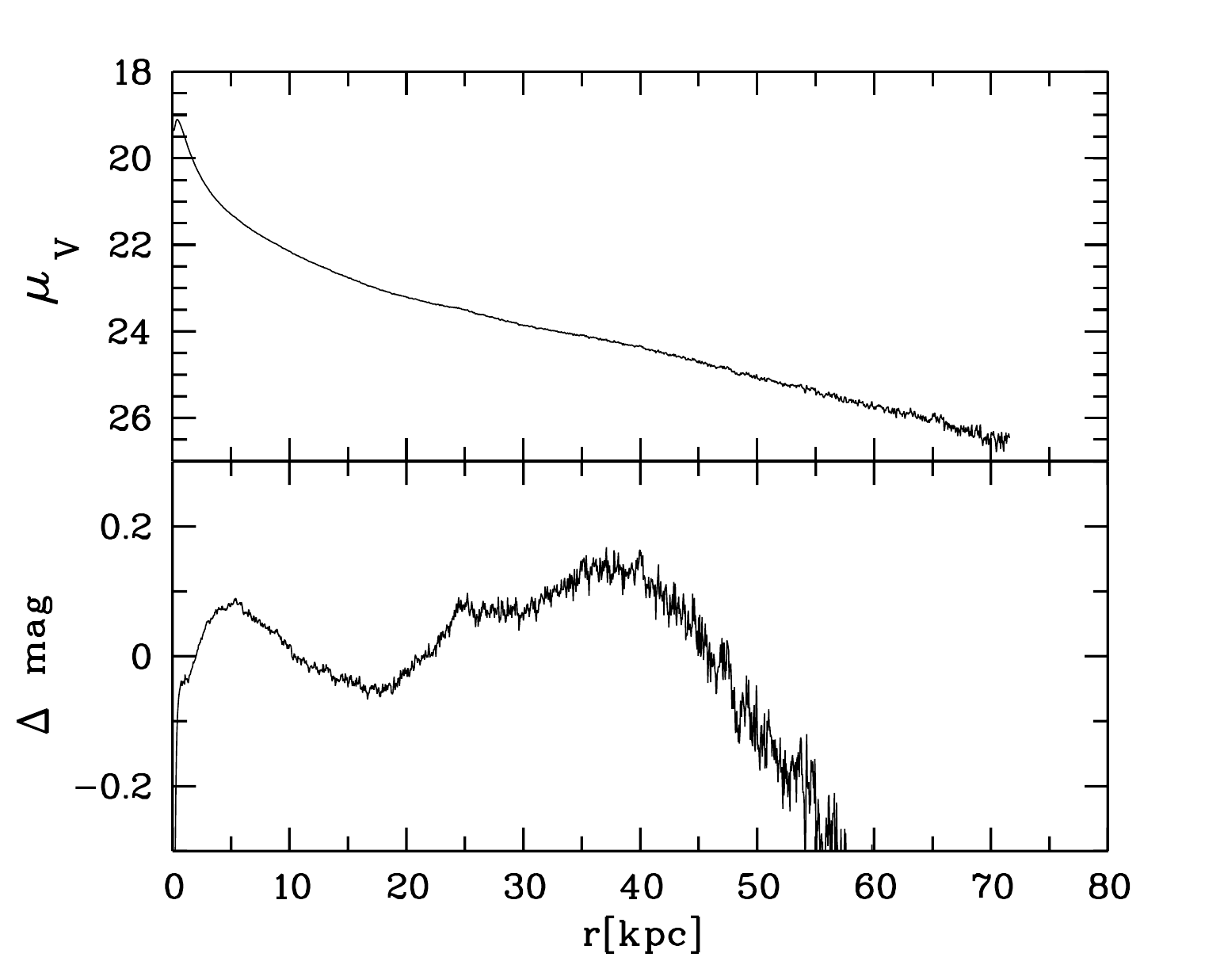}
\caption{Upper panel: Surface-brightness profile in the V-band.  Lower panel: residuals in mag between our measurements (upper panel) and
the photometric model in the sense: observations$-$model.}
\label{fig:light_profile}
\end{center}
\end{figure}

Fig.\ref{fig:light_profile} shows in its upper panel the measured surface brightness,  and in its lower panel
the difference of our photometric model with the measurements in the sense measurements-model.

 Out to 50 kpc,
the representation is excellent, the largest residuals being 0.1 mag. Then the model becomes gradually fainter. However, that
happens at such faint magnitudes that neither the deprojection nor the
cumulative luminosity are significantly affected. Furthermore, the accuracy of the photometry may be questioned.  While the low central surface brightness of NGC 3311 and its extended light
profile (besides its central position) are characteristics of a cD-galaxy, we cannot identify an inflection point \citep{kormendy89} which
would mark a transition between a  normal elliptical galaxy and a halo with a shallower light profile.  For example, the difference with
NGC 1399 in the Fornax cluster is the larger core of the NGC 3311 light profile. At large radii, both profiles decline
approximately as $R^{-2}$ \citep{dirsch03}.

\subsection{Globular clusters: dispersion and number density profile}
\label{sec:GCdispersion}
We  measured radial velocities for 118 bright globular clusters (GCs) in the absolute magnitude range $\mathrm -13.4 < M_V <  -10$. We use this label for simplicity although
the vast majority of these objects also can be
named ``Ultracompact Dwarfs'' (e.g. \citealt{mieske09}). Fig.\ref{fig:kinematics_V5} (left panel)  displays  the velocities versus  galactocentric distance. While for distances less
than 60-70 kpc the velocity dispersion seems to be statistically well defined, it is not the case for larger distances. It is clear that the
removal of the two objects near 5000 km/s with distances 100 kpc and 120 kpc would lower the dispersion considerably. 
However, a sudden drop of the dispersion beyond 100 kpc does not seem physical.  Moreover, the velocity dispersion
of the  cluster galaxies is about 800 km/s at 100 kpc  \citep{lokas06} which suggests  the outer GC dispersion to be of similar order.
We  consider the three open circles as obvious outliers and do not include them in the dispersion calculation. We think, however, that
these are  interesting objects, see the remarks in section\ref{sec:venti}. Here we only consider the full GC sample. Subsamples are discussed in Misgeld et al. 2011
(submitted to A\&A).

The upper panel of Fig.\ref{fig:models} shows the corresponding dispersion values and their uncertainties for 7 overlapping bins. Their widths  of 1.5-2 arcmin  assure   that
each bin contains about 30 objects (the two outermost bins have widths of 6 and 16 arcmin, respectively), enabling a solid statistical definition of the velocity dispersion (with the caveat given above for large radii).
 The velocity dispersions have been determined using the  estimator of \citet{pryor93}.
For a Jeans-analysis, we need  the number density profile of the GCs which we adopt from \citet{wehner08}, but their study reaches out to  3$\arcmin$ only and thus
must be extrapolated to larger radii.  For a convenient deprojection, we fit their  density profile to a Hubble-Reynolds law
(see e.g. \citealt{schuberth10} ; equation (11)) and obtain for the surface density 
\begin{equation}
N(r) = 3.86 (1+(r/8.2 \mathrm{kpc})^{-2})^{-0.7}
\end{equation}
with N(r) as  the number per kpc$^2$. 

 \subsection{The role of NGC 3309 and remarks on the validity of equilibrium}
 One might suspect a certain level of contamination of the cluster system of NGC 3311 by clusters from the
 nearby giant elliptical NGC 3309, which has a projected distance of only 100$\arcsec$ from NGC 3311. Already 
\citet{harris83} and \citet{mclaughlin95} estimated this contamination globally to be small considering the huge
cluster system of NGC 3311. \citet{wehner08} quote $16500 \pm 2000$ as  the total number of NGC 3311 clusters  in
comparison to
only $374 \pm  210$ in the case of 
NGC 3309 which means an abnormally low specific frequency of about 0.9 for NGC 3309.
This suggests that NGC 3309 might have lost part of its original cluster system to NGC 3311. The pair NGC 3311/NGC 3309
bears some resemblance to the central  pair NGC 1399/NGC 1404  in the Fornax cluster with NGC 1404 being
also underabundant in globular clusters \citep{richtler92} and clusters actually belonging to NGC 1404 indeed contaminate
the cluster system of NGC 1399 \citep{schuberth10}.  However, our sample of NGC 3311 clusters exclusively consists 
of very bright objects which are genuinely sparse in normal ellipticals, so we expect the contamination by NGC 3309 to be
even lower than the 2\% quoted by \citet{wehner08}. 

A further difference to NGC 1399/NGC 1404 is visible in the
X-ray isophotes. At radii larger than 8 kpc, the X-ray isophotes of NGC 1404 are distorted \citep{paolillo02} which
indicates the spatial proximity of NGC 1404 to NGC 1399, while no distortion is visible in the isophotes of NGC 3309
observed with Chandra \citep{yamasaki02} and XMM-Newton \citep{hayakawa06} These  observations support the assumption that NGC 3311 is at the center of
the cluster potential and NGC 3309 spatially at a larger distance from NGC 3311 than the projected one. Yamasaki et al. also point  out the good isothermality (the XMM-Newton temperature slightly decreases) and apparent equilibrium of the X-ray gas in
NGC 3311. Signs of an ongoing merger are at least not noticable. This is in line with the relative distance between  NGC 3309  
and NGC 3311 given by \citet{mieske05} using surface brightness fluctuations. NGC 3309 appears to be somewhat in the foreground.

The assumption of spherical equilibrium of  the  NGC 3311 cluster system and of Abell 1060 as a whole probably
is invalid to a certain degree. 
The notion of \citet{fitchett88} that the galaxy velocity distribution in Abell 1060 is flatter than a Gaussian, still holds
with the extended sample of  \citet{christlein03}. We did the excercise of performing a Shapiro-Wilk W-test for normality and found for
212 galaxies ($<$ 6000 km/s) within a radius of 600 kpc a p-value of 0.003 which indicates a non-Gaussian distribution
(note that this does not mean non-isotropy: in the galaxy cluster sample of \citealt{lokas06}, Abell 1060 is the most
isotropic cluster).  
On the other hand, our globular cluster sample
gives a p-value of 0.68, not contradicting a normal distribution.

\subsection{Stellar kinematics}
\label{sec:stellar}
 \begin{figure*}[]
\begin{center}
\includegraphics[width=18.0cm,angle=0]{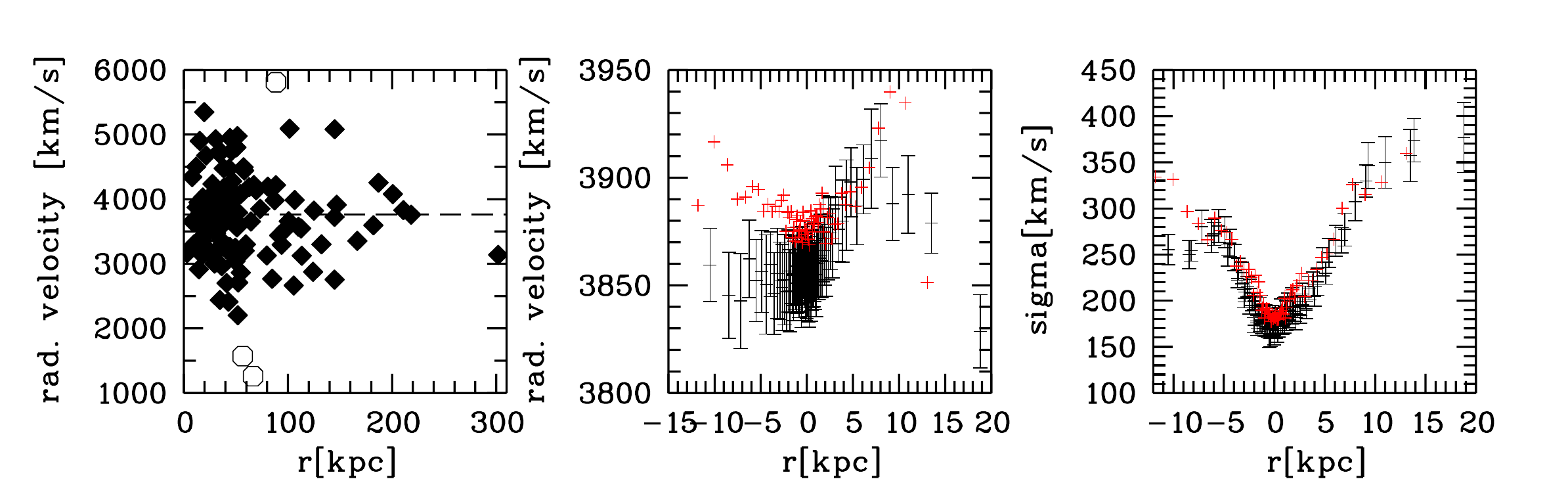}
\caption{Left: radial velocities of globular clusters (see Sec.\ref{sec:GCdispersion}). The horizontal dashed line is the mean heliocentric 
velocity of globular clusters inside 50 kpc which is 3765 km/s.
Middle: Radial velocity of NGC 3311 along the major axis (black crosses with error bars) and the minor axis (red crosses) (see
Sec.\ref{sec:stellar}). 
Right panel: Velocity dispersion of NGC 3311 along minor and major axis. }
\label{fig:kinematics_V5}
\end{center}
\end{figure*}

We analysed major axis (PA=28.9$^\circ$) and minor axis (PA=118.9$^\circ$) long slits. The NW-part of the minor axis is influenced by  NGC 3309 at radii larger than 10 kpc.
The  spectra are spatially rebinned in order to achieve a constant S/N
of 40 (major axis) and 30 (minor axis). 
Mean velocity and velocity dispersion 
were measured  employing the
 penalized pixel fitting (pPXF) method by
\citet{cappellari04}. In the pPXF analysis,  stellar templates from the 
  MILES
spectral library \citep{sanchez06} are convolved with a Gaussian model (plus higher order corrections) of the line-of-sight
velocity distribution until the $\chi^2$-deviations become minimal.

Fig.\ref{fig:kinematics_V5} (middle and right panel) shows the results.
 Radial distances are given in kpc with the negative values referring
to the SW-part of the major axis and to the NW-part of the minor axis. The radial velocity curve (middle) exhibits a slightly asymmetric shape in both axes, perhaps
indicating a small-scale complex velocity field.
The rise of the velocity dispersion (right panel) from a quite low central value of about 180 km/s to 350 km/s at 20 kpc 
(already shown in  \citealt{hau04})  resembles NGC 6166 in Abell 2199 (\citealt{carter99}, \citealt{kelson02})  where the dispersion rises to about
600  km/s at 60 kpc. 

The velocity dispersion along the major axis shows  some asymmetry as well in the sense that the SW part rises slightly faster than
the NE part, also seen in the minor axis. 
 \citet{loubser08} also show velocity dispersions for NGC 3311 with a radial limit of 15$\arcsec$(3.5 kpc). 
 Our measurements agree with their values in the common inner region. Recently, \citet{venti10} presented stellar kinematics of NGC 3311 out to
 30 kpc. Their interesting interpretation of  NGC 3311 is that of a small galaxy  embedded in a cluster-bound stellar halo,
 the transition occurring between 4 and 12 kpc. We plot their outer six bins  in the two lower panels of Fig.\ref{fig:models}.  Their two inner values (from Gemini/GMOS) 
 are compatible with our measurements while their intermediate values (from VLT/FORS2) are considerably higher. 
 Their outermost value  is again compatible with our model prediction (see \ref{sec:models}). Our major axis slit crosses their FORS2 slit
 in the NE-quadrant at a radius of about 50$\arcsec$(11.6 kpc). Large azimuthal differences over scales of 10$\arcsec$-
 20$\arcsec$ would be very surprising. 
  Furthermore, the comparison of the major and minor axes rather supports sphericity.  This
 issue remains open until a full map of kinematical data becomes available.
 


\subsection{X-ray gas mass}
On the galaxy cluster scale, the dominant baryonic component is the hot X-ray plasma. From various
X-ray studies of Abell 1060, we adopt the most recent XMM-Newton work by \citet{hayakawa06} and neglect
that their adopted distance is  slightly smaller (45.6 Mpc). 
One  finds the gas mass by numerical integration over a beta-profile for the gas density 
\begin{equation}
M_{gas} = 4 \pi n_0  \mu m_p \int_0^r (1+(z/r_c)^2)^{-\frac{3}{2} \beta} z^2 dz
\end{equation} 
where $n_0$= 11.7 cm$^{-3}$ is the central electron density, $\mu = 0.6$ the mean molecular weight, $m_p$ the proton mass, $r_c = 102$ kpc the scale radius, and $\beta = 0.69$.

In the isothermal case with kT= 3.3 KeV,
the dynamical mass is then given by e.g. \citet{grego01} 
\begin{equation}
\label{eq:massdyn}
M_{dyn} (r) = 2.5~10^8 \frac{r^3}{r_c^2 + r^2}[M_\odot/\mathrm{pc^3}] .
\end{equation}

We note, that this mass profile is not a good description for small radii: within 25 kpc,  the stellar mass
alone is larger than the dynamical mass. 
The notion of Hayakawa et al. that the central density profile is cuspy, therefore must be seen with caution.
 We find good agreement with their  gas mass profile. However, the gas mass has only a minor  influence on the dynamics,
 but we include
 it as the main baryonic component on larger scales.  The dynamical mass is used in section \ref{sec:massprofiles}.

\subsection{Comparison of globular cluster velocities with planetary nebulae}
\label{sec:venti}
During the revision process of this paper, \citet{venti11} published
radial velocities of a sample of 56 planetary nebulae (PNe) around NGC 3311. It is therefore
appropriate to make a brief comparison of PNe and GCs.

\begin{figure}[h]
\begin{center}
\includegraphics[width=0.50\textwidth]{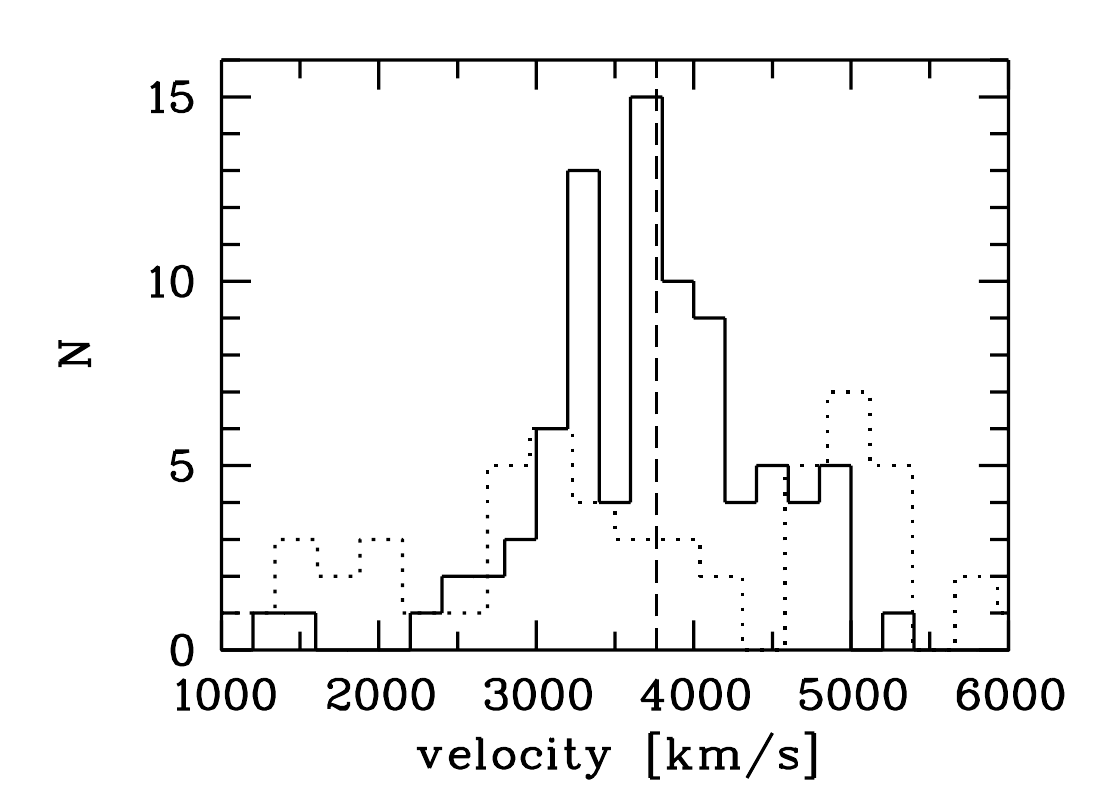}
\caption{A comparison of the velocity distributions of globular clusters (GCs; solid histogram)  and  planetary nebulae (PNe; dashed histogram) from \citet{venti11}.  The vertical dashed line
at 3765 km/s indicates the mean of all GC velocities. The GCs have
projected galactocentric distances smaller than 65 kpc in order to match the radial range of PNe. There are striking differences whose
origin is not yet understood.
}
\label{fig:venti}
\end{center}
\end{figure}

Fig.\ref{fig:venti} compares the velocity histograms of PNe and GCs inside
a radius of 65 kpc which is set by the radial extent of the PNe sample.
The first remark is that the PNe sample suffers even more from
small number statistics than the GC sample. Any conclusion has to be
confirmed with a larger sample. It is, however, striking that the velocity
distribution lacks the peak
around the velocity of NGC 3311 which the GC sample prominently shows. \citet{venti11} suggest (along with the speculation that NGC 3311 is intrinsically devoid of
PNe)  that ram pressure stripping of the PN-shells in the dense parts
of the hot X-ray plasma might be responsible by shortening the lifetimes
of PNe. In this case the PNe are sampled outside a cavity around NGC 3311.
The questions remain why such effect should remove the velocity peak and why
this is neither visible in NGC 1399 \citep{mcneil10} nor in M87 \citep{doherty09} where the gas densities are of comparable order.
One could add the speculation that the PNe stem from different stellar parent
populations, perhaps partly of intermediate-age reminiscent of earlier
infall of (dwarf?)-galaxies.

The peak at 3300 km/s in the GC distribution is within the present sample
not trustworthy. Only a few more objects with about 3500 km/s would suffice to
erase it. However, there is a similarity with PNe concerning  GCs at larger radii than 100 kpc.
As visible in Fig.\ref{fig:kinematics_V5}, there seems to be a bias
towards velocities below the systemic velocity.  Again: whether this a sample size
effect or not, must be investigated with a larger sample.

Interesting is the occurrence of PNe with velocities smaller than 1500 km/s in which
range also two GCs are found (1293 km/s and 1570 km/s, Misgeld et al. 2011, submitted to A\&A ). These are  deviations of partly more than 3$\sigma$ for
a Gaussian distribution centered on NGC 3311, corresponding to a probability of 0.1\% or less. For the GCs it is already  improbable to
have objects at this velocity, for the PNe due their smaller sample and three
objects even more so. The much larger samples of galaxies \citep{christlein03}
and dwarf galaxies \citep{misgeld08} do not contain velocities below 2000 km/s.

The question is therefore whether these low velocities are recession velocities
rather than Doppler-velocities. While the existence of stellar populations
between galaxy clusters would be intriguing, one has to wait for future surveys.  
However, very high peculiar velocities are principally possible. We come back to this
issue in section \ref{sec:orbits}.                                                                         

\section{Dynamical models}
\label{sec:models}

\begin{figure}[h!]
\begin{center}
\includegraphics[width=0.45\textwidth,angle=0]{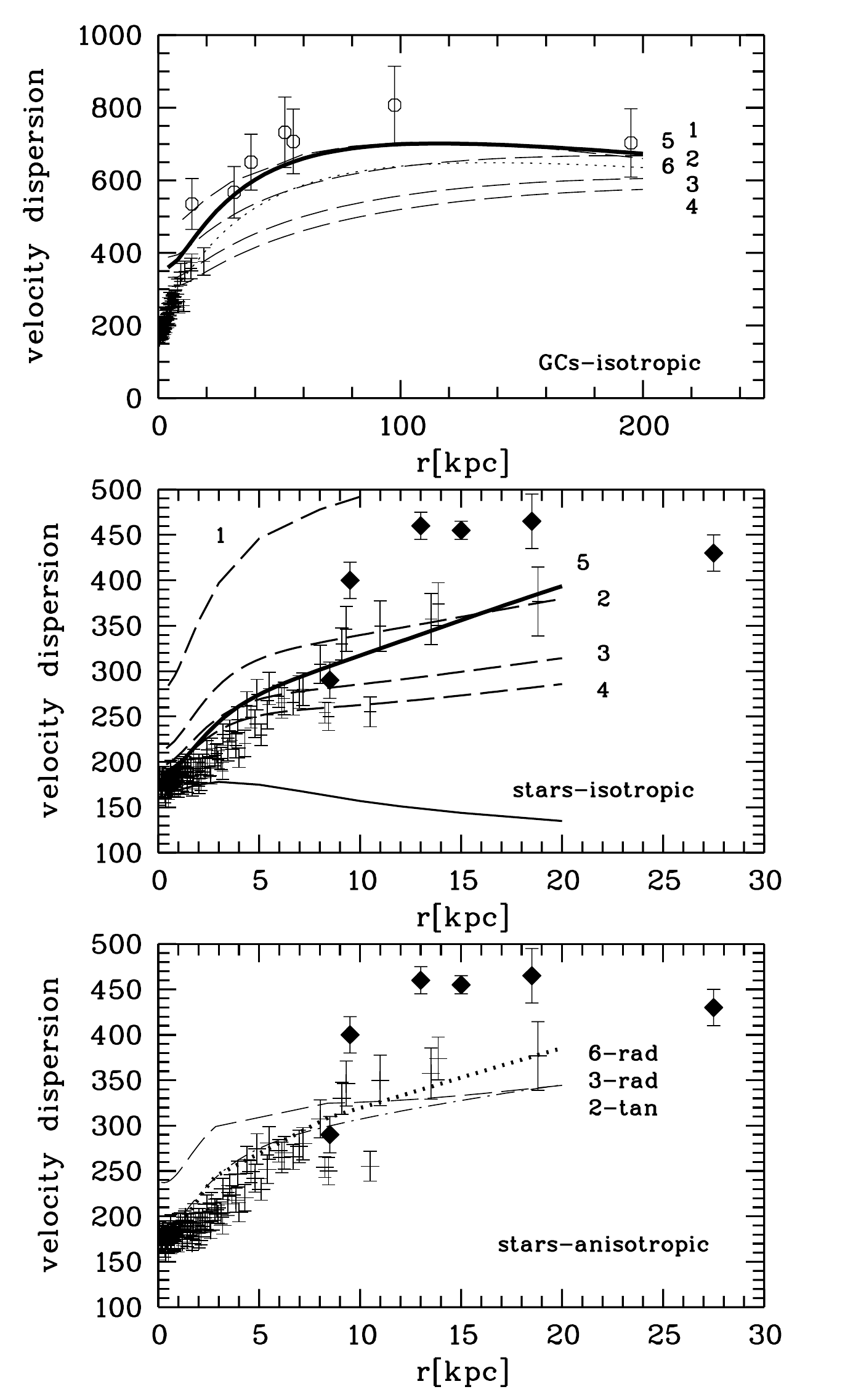}
\caption{{\bf Upper panel}: the extended regime of the globular clusters and their velocity dispersions. Open circles denote
globular clusters, crosses (which merge together) denote stars. The kinematical data of \citet{venti10} are plotted in the lower two panels
See section \ref{sec:stellar} for more remarks.
 Six isotropic models (stellar mass+gas mass+dark halo),
based on different halo shapes,  are indicated. The halo properties are listed in Table \ref{tab:models}. Thick solid line (halo 5): a cored halo (eq. \ref{eq:burkert}). Dashed lines: NFW halos (eq.\ref{eq:NFW}).  Numbering is the
sequence of vertical order. Halo 1 is  the NFW halo quoted by
\citet{lokas06} from the analysis of galaxy velocities in Abell 1060 (eq. 6) (almost coinciding with halo 5).  Dotted line (almost coinciding with
halo 2) : halo 6, which is a cored Burkert halo designed to represent the stars under a radial anisotropy.
 {\bf Middle panel}: the regime of stellar kinematics with isotropic models using  the same numbering.
 The kinematical data of \citet{venti10} are plotted as black diamonds (also in the lower panel). Halo 1 deviates strongly, but is not cosmological according to \citet{maccio08}.
The lower solid line without number is a model without dark matter for comparison.
{\bf Lower panel}: these three models demonstrate the effect of  anisotropy  of the stellar orbits. Halo 3: radial
anisotropy according to \citet{hansen06}. Halo 2: constant tangential anisotropy of -0.5. Halo 6 is a Burkert halo
with radial anisotropy.     See sections \ref{sec:models} and \ref{sec:stellar} for details.
 }
\label{fig:models}
\end{center}
\end{figure}

We  present dynamical models which are solutions of the non-rotating spherical Jeans-equation (see \citealt{mamon05}
 for a collection of the
relevant formulae). They can be considered 
good approximations to  spherical stellar systems, assuming equilibrium. We further assume 
that NGC 3311 is at rest at the center of the cluster potential.  Given the data limitations,  we regard
our results as a first approximation to  the mass distribution which will be refined later both observationally and theoretically.

With the stellar mass component, the gaseous component and an assumed dark halo, for which we try out various
representations, we model the
projected velocity dispersions versus projected radius, using the formulae quoted in \citet{mamon05}.
We require a dark halo, which simultaneously
can account for the dispersion of globular clusters and  stars.

The upper panel in Fig.\ref{fig:models} shows the dispersion of GCs as described in section \ref{sec:GCdispersion}. 
Various models are indicated.  All use $M/L_V=6$ for the stellar population which is needed to reproduce
the inner stellar velocity dispersion, and which also is close to the expected value for an old metal-rich population
such as NGC 3311 (Spolaor et al. 2011, submitted to MNRAS). The middle panel shows the velocity dispersion
of the stellar population where the same models are indicated. These models assume isotropy for the stellar
distribution of orbits.  
To demonstrate the need for dark matter, the lower solid line is the expected dispersion without a dark halo.
The lower panel shows for selected halos the effect of radial and tangential anisotropies. The kinematical data
of \citet{venti10} are plotted In the lower two panels.

In the following we discuss the models individually.

\subsection{NFW halos}
The dark halo of Abell 1060 has been previously investigated by \citet{lokas06} by a Jeans-analysis, using published galaxy velocities.
They use NFW-halos (e.g. \citealt{navarro04}) which are analytically represented by 

\begin{equation}
\rho(r) = \frac{\rho_s}{\frac{r}{r_s} (1+\frac{r}{r_s})}
\label{eq:nfwprofile}
\end{equation}
 where $\rho_s$ and $r_s$ are the characteristic density and
radius, respectively. 
They have the  cumulative mass 
\begin{equation}  
M_{dark} = 4 \pi \rho_s r_s^3  \left(\ln((r_s+r)/r_s) - r/(r_s+r) \right)
\label{eq:NFW}
\end{equation}
 It is customary to define a virial radius $r_{vir}$ (which includes the virial mass $M_{vir}$) inside which the mean density is by a certain factor higher
than the critical density of the Universe. This factor is mostly set to 200. A further parameter is the ''concentration'' $c= r_{vir}/r_s$.
Simulations agree in that there is a relation between $M_{vir}$ and c, albeit with some scatter. \citep{maccio08} quote for
that relation (we adopt their eq.10 for relaxed halos and set h=1 for convenience)

\begin{equation}
\log~ c = 0.83 - 0.098 (\log (M_{vir}/10^{12}).
\label{eq:maccio}
\end{equation}

In Fig.\ref{fig:models}, the upper panel refers to the regime of globular clusters, the middle panel to the regime of
stars, showing isotropic models, the lower panel as well to the stars, but with a radial and a tangential model
indicated. The numbers always indicate the same halos.   For the halo properties, see Table \ref{tab:models}.

 Halo 1 in Fig.\ref{fig:models}  shows  the best fitting NFW halo of \citet{lokas06}
with $\rho_s = 0.0072 M_\odot/pc^3$ and $r_s = 140$ kpc  $M_{vir}$ is $3.87\times10^{14} M_\odot$ and c=10.9 
(note that these values use a factor 200 for defining the virial mass, while \citet{lokas06} use 101.9 and therefore quote c= 14). This halo marginally fits the
globular clusters, but definitely not the stars. It has a higher  concentration than a cosmological halo in the sense of eq.\ref{eq:maccio},
from where one would expect c=3.4. However, \citet{lokas06} give a large uncertainty range for their c-value, which expresses the fact that
the galaxies do not strongly constrain the halo shape.

We therefore try out halos which approximately fulfill the Maccio et al.-relation and which do not deviate too strongly
from the mass quoted by \citet{lokas06}.  Halo 3 falls onto this relation while halo 2 and 4 are approximately 1-sigma deviations.

Halo 3 and 4 are neither good fits for the globular cluster under isotropy (upper panel) nor do they fit the stars 
beyond about 8 kpc (middle panel). In order to make halo 3 a better representation, one could introduce a 
stellar radial anisotropy.

Modest radial anisotropies are common in elliptical galaxies and are also expected
from simulations of mergers (\citealt{hansen06}, \citealt{mamon06}). A radial anisotropy  elevates the velocity dispersions, most strongly in the central regions.  The prescription of  \citet{hansen06} minimizes 
the central elevation

\begin{equation}
\label{eq:radial}
\beta = -0.1 - 0.2 \frac{d~ \ln \rho(r)}{d~ \ln r} r
\end{equation}
where the logarithmic slope of the baryonic mass profile enters. $\beta$ is the anisotropy parameter. Our mass profile gives a somewhat complicated
behavior, but a very good
approximation is the form $\beta = 0.5 r/(r_a+r)$ with $r_a$=3 kpc which corresponds to eq. 60 of \citet{mamon05} and
permits us to employ their analytical solutions. 

This radial anisotropy for halo 3 is displayed in the lower panel. With some more fine-tuning, one could probably
reach a good representation of the stellar kinematics. The problem with globular clusters of course remains.

Halo 2 is so far the most promising, although not really satisfactory. The effect of a slight, radially constant tangential anisotropy of $\beta = -0.5$ is shown in the lower
panel.  One needs, however, a higher mass and/or a radial anisotropy of the globular clusters. To boost the
velocity dispersion at, say, 100 kpc from 600 km/s to 750 km/s, one needs approximately a factor 1.6 in mass (which
may be difficult to justify),
but then a stronger tangential isotropy together with some fine-tuning is needed. We remark, however, that  
 tangential anisotropies  in elliptical galaxies seem to be rare but see NGC 1407 for an example \citep{romanowsky09}.

\subsection{A cored halo}

A more straight-forward  representation can be achieved by considering a halo with a core instead of a cusp.
One  could think of 
the simple  density profile $\rho(r) = \rho_0/(1+(r/r_c)^2)$, $\rho_0$ being the central density and $r_c$ a scale radius.
However, this profile, although it permits a good representation for small radii, corresponds to very high masses for large radii, which deviate strongly from the X-ray mass profile. We therefore use the "Burkert profile" \citep{burkert95} which
has been introduced for the dark matter profiles of spiral dwarf galaxies

\begin{equation}
 \rho(r) = \frac{\rho_0}{(1+r/r_c)(1+(r/r_c)^2)}
 \end{equation}
 
 The cumulative mass is
 \begin{equation} 
 M_{dark} = 4\pi \rho_0 r_c^3  \left[\frac{1}{2} ln(1+r/r_c) +  \frac{1}{4} ln(1+(r/r_c)^2)  - \frac{1}{2} \arctan{(r/r_c)} \right]
 \label{eq:burkert}
 \end{equation}
 
 This halo is shown in Fig.\ref{fig:models} as the thick solid line (halo 5). Such halo seems to provide the best representation.
 
 To illustrate how the radial anisotropy of eq.\ref{eq:radial} affects this kind of halo, we plot halo 6 in the lower 
 panel (solid thick line) which fits  the stars quite well.  However, we had to decrease the stellar M/L to 3 in order
 not to boost the central velocity dispersion. This low M/L-value is not supported by Spolaor et al. (submitted to MNRAS).
 Also here, a radial bias needs a lower halo mass
 which in turn causes difficulties to fit the globular clusters.





\begin{table}[h]
\caption{Properties of dark matter halos in Fig.\ref{fig:models}. Halos 1- 4 are of the NFW type (eq.\ref{eq:NFW}),
halo 5 is a cored Burkert halo (eq.\ref{eq:burkert}).  Column 1: ID, column2: characteristic density (central density in case
of halo 5), column 3: scale radius, column 4: virial mass (for the Burkert halos the mass within 1 Mpc), column 5: concentration, column 6: type.}
\resizebox{9cm}{!}{
\begin{tabular}{cccccl}
ID &  $\rho_s [M_\odot/pc^3] $ & $r_s [kpc]$ & $M_{vir} [10^{14} M_\odot]$ & c & type \\
\hline
1  &   0.0072 & 140 &  3.87 & 10.9 & NFW \\
2 & 0.0013 & 290 & 3.97 & 5.3 & NFW \\
3  & 0.00051 & 440 & 3.94 & 3.94 & NFW\\
4  &  0.0003 & 570 &  4.05 & 2.7 & NFW\\
5 & 0.02 & 75 & 2.0 (1 Mpc) & - & Burkert halo (isotropic)\\ 
6 & 0.012 & 90 & 1.9  (1 Mpc) & - & Burkert halo (radial)\\ 
\hline
\label{tab:models}

\end{tabular}
}
\end{table}%

\subsection{Comparison with the X-ray dynamical mass and with the halos of other nearby
central galaxies}
\label{sec:massprofiles}

\begin{figure}[h!]
\begin{center}
\includegraphics[width=0.4\textwidth,angle=0]{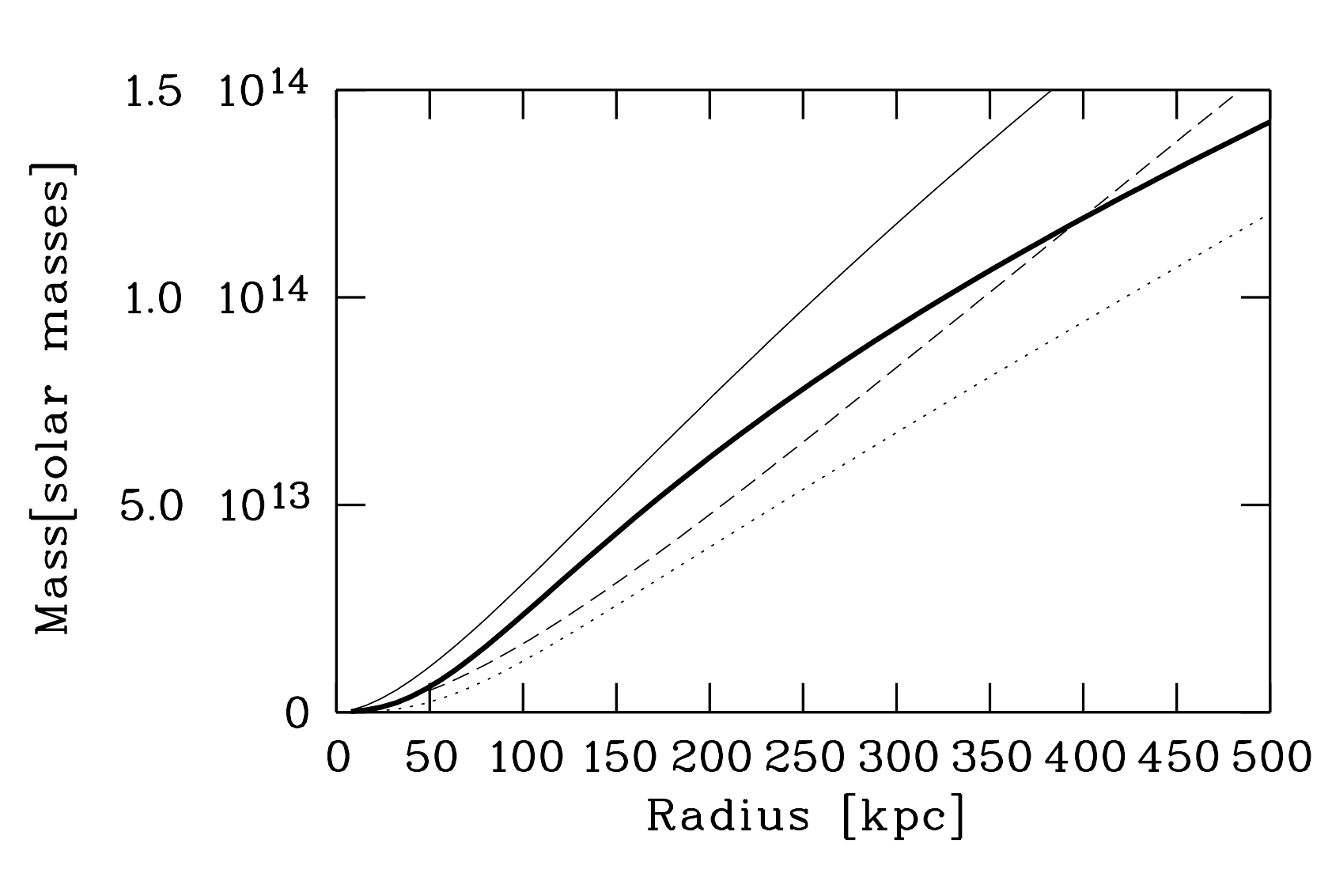}
\caption{Mass profiles for some halos from Table \ref{tab:models} compared with the X-ray based mass from eq.\ref{eq:massdyn} (dotted line). Solid thick line: Burkert halo (halo 5), thin solid line: halo 1, dashed line: halo 2}.
\label{fig:massprofiles}
\end{center}
\end{figure}

It is interesting to compare the various halos with the X-ray based mass (\citealt{hayakawa06}; eq.\ref{eq:massdyn}) on larger scales.

Fig.\ref{fig:massprofiles} 
shows total mass profiles (stellar mass of NGC 3311 + gaseous mass+dark mass) for halos from Table  \ref{tab:models}.  The thick solid line is the Burkert-halo, the thin solid line is halo 1, and the dashed line is halo 2. With the
exception of halo 1, they are in reasonable agreement at large distances with the X-ray based  mass. We recall that halo 1 is
not well defined, so its deviation might not be alarming.  We moreover remark that the X-ray mass also is an extrapolation with the
outermost measurement at about 170 kpc.  It is amazing that the extrapolation out of  a small volume of 20 kpc radius 
still results in sensible masses on the scale of  several hundreds kpc. This raises
an oddity: halo 2 represents a ''cosmological'' cluster-wide halo, but there is no room anymore for a galaxy-wide halo.

Thus the central dark matter  in NGC 3311 ``knows'' about its location within a galaxy cluster. While this is plausible with view
on pure dark matter simulations, one should conclude that baryons were not strongly involved in shaping the  inner dark matter profile.

A transition between a galaxy halo and a cluster halo has been claimed to be detected in the X-ray profile of NGC 1399,
the central galaxy of the Fornax cluster \citep{ikebe96}.  However,  this feature could not be found in the kinematics of
the GC system  \citep{schuberth10}.  The best fitting dark halos (see Table 6 of Schuberth et al.) are those which
 use only the metal-rich GCs. These halos have small core radii and clearly are not cosmological.  The metal-poor
 GCs show a more erratic behaviour and indicate halos which are marginally cosmological, but fit worse.
 

On the other hand, the dark halo of M87, the central galaxy of the Virgo cluster, which is advocated by \citet{cote01} (NFW halo, their eq.11), is similar to
our halo 4 with
  $\rho_s = 0.0003 M_\odot/pc^3$ , $r_s = 560$ kpc, and c= 2.8.

In conclusion, there is not yet a uniform picture for the halos of nearby central galaxies.

\subsection{Extreme velocities}
\label{sec:orbits}
Some GCs and PNe show radial velocities with offsets of about 2700 km/s with respect to the mean cluster velocity. Are such
orbits consistent with the halo masses? 
We consider the most favorable case, namely that these velocities are space velocities near their perigalactica. We adopt halo  2,
neglecting stellar and gas masses. Our GC with the lowest velocity (1259 km/s) has a projected distance of 66 kpc from NGC 3311.
We use an orbit program kindly provided by M. Fellhauer (private communication) to calculate the apogalactic distance if the perigalactic
distance is 66 kpc and the perigalactic velocity is 2465 km/s. The result is an apogalactic distance of 1237 kpc and an orbital
period of 2.8 Gyr.  These orbits are thus possible, but since it is  improbable to pick up 
an object near its pericenter while its velocity vector points towards the observer, one may expect a larger population of these
`''intra-cluster" objects with less inclined orbits. These have also been found among GCs in the Fornax  \citep{schuberth10,schuberth08} and Coma   \citep{peng11} galaxy clusters. PNe associated with the intra-cluster light in Hydra I have been discussed in
\citet{venti08}. This is in line with studies finding the contribution of intra-cluster stellar populations 
to be considerable, e.g. \citet{gonzalez07}.

\section{Conclusions} 
The core-cusp controversy regarding the inner shape of dark matter halos in galaxies seems to be far from being
resolved.    Particularly in low surface brightness galaxies,  cuspy halos which are predicted by cosmological simulations,
do not find support in the observational data (e.g. \citealt{deblok10}, but see \citealt{goerdt10} for a mechanism of how to transform
a cuspy dark matter profile into a cored profile). Therefore, it is of fundamental interest to investigate the more massive
halos of galaxy clusters. 
NGC 3311, the central galaxy in the Hydra I cluster (Abell 1060), appears as an interesting test object due to its low surface brightness
which makes dark matter effects better visible.

We use the velocity dispersion profile of the stellar population of NGC 3311  together with 118
radial velocities of bright globular clusters for an attempt to constrain the shape of the dark matter
halo. 
With simplifying assumptions (equilibrium, sphericity,  NGC 3311 at rest),  we find that  cuspy dark matter halos of the NFW type, which obey the relation between concentration and virial mass found in cosmological simulations and  which explain the globular cluster velocity dispersion,   do not satisfactorily account for the smooth rise of the stellar velocity dispersion at small radii under isotropy.  Halos with a core better reproduce the stellar and globular cluster
kinematics observed for NGC 3311.

A radial anisotropy, which is frequently observed in elliptical galaxies, aggravates the problem for any kind of halo in that
the projected velocity dispersions are boosted and the halo mass becomes too low to account for the globular clusters.  Tangential anisotropies (best a fine-tuned radial behaviour) could
perhaps provide a solution, but they never have been found observationally neither in dynamical models of galaxies.

 With the present analysis, one cannot claim  that  a cuspy halo of NGC 3311 is ruled out, but a cored dark matter halo seems to be favored.

 Our assumptions  may be questioned. More sophisticated dynamical modelling on the basis of an extended data set
 (many more velocities, a better knowledge of the structural parameters of the globular cluster system, and a precise two-dimensional velocity field of the NGC 3311 galaxy light) may find slightly different halos. However, the finding of a core would
 be  in line with other work pointing at shallow dark matter profiles in galaxy
 clusters (see introduction). Any resolution of the conflict with the standard knowledge of dark matter must work on a very large mass scale from dwarf galaxies 
 (e.g. \citealt{gentile07}) to galaxy clusters.  A comparison with other nearby central galaxies does not yet reveal a uniform picture.   

The potential of NGC 3311 and other central galaxies with low surface brightness to trace the dark cores of galaxy
clusters awaits exciting future work.

\begin{acknowledgements}
We thank an anonymous referee for a thoughtful report and particularly for emphasizing the role
of the concentration-mass relation of cosmological halos.  We thank Michael Fellhauer for providing
his orbit program.
TR acknowledges financial support from the Chilean Center for Astrophysics,
FONDAP Nr. 15010003,  from FONDECYT project Nr. 1100620, and
from the BASAL Centro de Astrofisica y Tecnologias
Afines (CATA) PFB-06/2007. IM acknowledges support through DFG grant BE1091/13-1.
 AJR was supported by National Science Foundation
grants AST-0808099 and AST-0909237.
\end{acknowledgements}
\bibliographystyle{aa}
\bibliography{N3311}

\begin{thebibliography}{52}
\expandafter\ifx\csname natexlab\endcsname\relax\def\natexlab#1{#1}\fi

\bibitem[{{Bullock} {et~al.}(2001){Bullock}, {Kolatt}, {Sigad}, {Somerville},
  {Kravtsov}, {Klypin}, {Primack}, \& {Dekel}}]{bullock01}
{Bullock}, J.~S., {Kolatt}, T.~S., {Sigad}, Y., {et~al.} 2001, \mnras, 321, 559

\bibitem[{{Burkert}(1995)}]{burkert95}
{Burkert}, A. 1995, \apjl, 447, L25+

\bibitem[{{Cappellari} \& {Emsellem}(2004)}]{cappellari04}
{Cappellari}, M. \& {Emsellem}, E. 2004, \pasp, 116, 138

\bibitem[{{Carter} {et~al.}(1999){Carter}, {Bridges}, \& {Hau}}]{carter99}
{Carter}, D., {Bridges}, T.~J., \& {Hau}, G.~K.~T. 1999, \mnras, 307, 131

\bibitem[{{Christlein} \& {Zabludoff}(2003)}]{christlein03}
{Christlein}, D. \& {Zabludoff}, A.~I. 2003, \apj, 591, 764

\bibitem[{{C{\^o}t{\'e}} {et~al.}(2001){C{\^o}t{\'e}}, {McLaughlin}, {Hanes},
  {Bridges}, {Geisler}, {Merritt}, {Hesser}, {Harris}, \& {Lee}}]{cote01}
{C{\^o}t{\'e}}, P., {McLaughlin}, D.~E., {Hanes}, D.~A., {et~al.} 2001, \apj,
  559, 828

\bibitem[{{de Blok}(2010)}]{deblok10}
{de Blok}, W.~J.~G. 2010, Advances in Astronomy, 2010

\bibitem[{{Diemand} {et~al.}(2005){Diemand}, {Zemp}, {Moore}, {Stadel}, \&
  {Carollo}}]{diemand05}
{Diemand}, J., {Zemp}, M., {Moore}, B., {Stadel}, J., \& {Carollo}, C.~M. 2005,
  \mnras, 364, 665

\bibitem[{{Dirsch} {et~al.}(2003){Dirsch}, {Richtler}, {Geisler}, {Forte},
  {Bassino}, \& {Gieren}}]{dirsch03}
{Dirsch}, B., {Richtler}, T., {Geisler}, D., {et~al.} 2003, \aj, 125, 1908

\bibitem[{{Doherty} {et~al.}(2009){Doherty}, {Arnaboldi}, {Das}, {Gerhard},
  {Aguerri}, {Ciardullo}, {Feldmeier}, {Freeman}, {Jacoby}, \&
  {Murante}}]{doherty09}
{Doherty}, M., {Arnaboldi}, M., {Das}, P., {et~al.} 2009, \aap, 502, 771

\bibitem[{{Fitchett} \& {Merritt}(1988)}]{fitchett88}
{Fitchett}, M. \& {Merritt}, D. 1988, \apj, 335, 18

\bibitem[{{Gentile} {et~al.}(2007){Gentile}, {Salucci}, {Klein}, \&
  {Granato}}]{gentile07}
{Gentile}, G., {Salucci}, P., {Klein}, U., \& {Granato}, G.~L. 2007, \mnras,
  375, 199

\bibitem[{{Goerdt} {et~al.}(2010){Goerdt}, {Moore}, {Read}, \&
  {Stadel}}]{goerdt10}
{Goerdt}, T., {Moore}, B., {Read}, J.~I., \& {Stadel}, J. 2010, \apj, 725, 1707

\bibitem[{{Gonzalez} {et~al.}(2007){Gonzalez}, {Zaritsky}, \&
  {Zabludoff}}]{gonzalez07}
{Gonzalez}, A.~H., {Zaritsky}, D., \& {Zabludoff}, A.~I. 2007, \apj, 666, 147

\bibitem[{{Grego} {et~al.}(2001){Grego}, {Carlstrom}, {Reese}, {Holder},
  {Holzapfel}, {Joy}, {Mohr}, \& {Patel}}]{grego01}
{Grego}, L., {Carlstrom}, J.~E., {Reese}, E.~D., {et~al.} 2001, \apj, 552, 2

\bibitem[{{Hansen} \& {Moore}(2006)}]{hansen06}
{Hansen}, S.~H. \& {Moore}, B. 2006, \na, 11, 333

\bibitem[{{Harris} {et~al.}(1983){Harris}, {Smith}, \& {Myra}}]{harris83}
{Harris}, W.~E., {Smith}, M.~G., \& {Myra}, E.~S. 1983, \apj, 272, 456

\bibitem[{{Hau} {et~al.}(2004){Hau}, {Hilker}, {Bridges}, {Carter}, {Dejonghe},
  {de Rijcke}, \& {Quintana}}]{hau04}
{Hau}, G.~K.~T., {Hilker}, M., {Bridges}, T., {et~al.} 2004, in IAU Colloq.
  195: Outskirts of Galaxy Clusters: Intense Life in the Suburbs, ed.
  {A.~Diaferio}, 491--495

\bibitem[{{Hayakawa} {et~al.}(2006){Hayakawa}, {Hoshino}, {Ishida}, {Furusho},
  {Yamasaki}, \& {Ohashi}}]{hayakawa06}
{Hayakawa}, A., {Hoshino}, A., {Ishida}, M., {et~al.} 2006, \pasj, 58, 695

\bibitem[{{Ikebe} {et~al.}(1996){Ikebe}, {Ezawa}, {Fukazawa}, {Hirayama},
  {Ishisaki}, {Kikuchi}, {Kubo}, {Makishima}, {Matsushita}, {Ohashi},
  {Takahashi}, \& {Tamura}}]{ikebe96}
{Ikebe}, Y., {Ezawa}, H., {Fukazawa}, Y., {et~al.} 1996, \nat, 379, 427

\bibitem[{{Kelson} {et~al.}(2002){Kelson}, {Zabludoff}, {Williams}, {Trager},
  {Mulchaey}, \& {Bolte}}]{kelson02}
{Kelson}, D.~D., {Zabludoff}, A.~I., {Williams}, K.~A., {et~al.} 2002, \apj,
  576, 720

\bibitem[{{Kormendy} \& {Djorgovski}(1989)}]{kormendy89}
{Kormendy}, J. \& {Djorgovski}, S. 1989, \araa, 27, 235

\bibitem[{{{\L}okas} {et~al.}(2006){{\L}okas}, {Wojtak}, {Gottl{\"o}ber},
  {Mamon}, \& {Prada}}]{lokas06}
{{\L}okas}, E.~L., {Wojtak}, R., {Gottl{\"o}ber}, S., {Mamon}, G.~A., \&
  {Prada}, F. 2006, \mnras, 367, 1463

\bibitem[{{Loubser} {et~al.}(2008){Loubser}, {Sansom},
  {S{\'a}nchez-Bl{\'a}zquez}, {Soechting}, \& {Bromage}}]{loubser08}
{Loubser}, S.~I., {Sansom}, A.~E., {S{\'a}nchez-Bl{\'a}zquez}, P., {Soechting},
  I.~K., \& {Bromage}, G.~E. 2008, \mnras, 391, 1009

\bibitem[{{Macci{\`o}} {et~al.}(2008){Macci{\`o}}, {Dutton}, \& {van den
  Bosch}}]{maccio08}
{Macci{\`o}}, A.~V., {Dutton}, A.~A., \& {van den Bosch}, F.~C. 2008, \mnras,
  391, 1940

\bibitem[{{Mamon} {et~al.}(2006){Mamon}, {{\L}okas}, {Dekel}, {Stoehr}, \&
  {Cox}}]{mamon06}
{Mamon}, G.~A., {{\L}okas}, E., {Dekel}, A., {Stoehr}, F., \& {Cox}, T.~J.
  2006, in EAS Publications Series, Vol.~20, EAS Publications Series, ed.
  {G.~A.~Mamon, F.~Combes, C.~Deffayet, \& B.~Fort}, 139--148

\bibitem[{{Mamon} \& {{\L}okas}(2005)}]{mamon05}
{Mamon}, G.~A. \& {{\L}okas}, E.~L. 2005, \mnras, 363, 705

\bibitem[{{McLaughlin} {et~al.}(1995){McLaughlin}, {Secker}, {Harris}, \&
  {Geisler}}]{mclaughlin95}
{McLaughlin}, D.~E., {Secker}, J., {Harris}, W.~E., \& {Geisler}, D. 1995, \aj,
  109, 1033

\bibitem[{{McNeil} {et~al.}(2010){McNeil}, {Arnaboldi}, {Freeman}, {Gerhard},
  {Coccato}, \& {Das}}]{mcneil10}
{McNeil}, E.~K., {Arnaboldi}, M., {Freeman}, K.~C., {et~al.} 2010, \aap, 518,
  A44+

\bibitem[{{Mieske} {et~al.}(2005){Mieske}, {Hilker}, \& {Infante}}]{mieske05}
{Mieske}, S., {Hilker}, M., \& {Infante}, L. 2005, \aap, 438, 103

\bibitem[{{Mieske} {et~al.}(2009){Mieske}, {Hilker}, {Misgeld}, {Jord{\'a}n},
  {Infante}, \& {Kissler-Patig}}]{mieske09}
{Mieske}, S., {Hilker}, M., {Misgeld}, I., {et~al.} 2009, \aap, 498, 705

\bibitem[{{Misgeld} {et~al.}(2008){Misgeld}, {Mieske}, \& {Hilker}}]{misgeld08}
{Misgeld}, I., {Mieske}, S., \& {Hilker}, M. 2008, \aap, 486, 697

\bibitem[{{Navarro} {et~al.}(2004){Navarro}, {Hayashi}, {Power}, {Jenkins},
  {Frenk}, {White}, {Springel}, {Stadel}, \& {Quinn}}]{navarro04}
{Navarro}, J.~F., {Hayashi}, E., {Power}, C., {et~al.} 2004, \mnras, 349, 1039

\bibitem[{{Newman} {et~al.}(2011){Newman}, {Treu}, {Ellis}, \&
  {Sand}}]{newman11}
{Newman}, A.~B., {Treu}, T., {Ellis}, R.~S., \& {Sand}, D.~J. 2011, \apjl, 728,
  L39+

\bibitem[{{Newman} {et~al.}(2009){Newman}, {Treu}, {Ellis}, {Sand}, {Richard},
  {Marshall}, {Capak}, \& {Miyazaki}}]{newman09}
{Newman}, A.~B., {Treu}, T., {Ellis}, R.~S., {et~al.} 2009, \apj, 706, 1078

\bibitem[{{Paolillo} {et~al.}(2002){Paolillo}, {Fabbiano}, {Peres}, \&
  {Kim}}]{paolillo02}
{Paolillo}, M., {Fabbiano}, G., {Peres}, G., \& {Kim}, D. 2002, \apj, 565, 883

\bibitem[{{Peng} {et~al.}(2011){Peng}, {Ferguson}, {Goudfrooij}, {Hammer},
  {Lucey}, {Marzke}, {Puzia}, {Carter}, {Balcells}, {Bridges}, {Chiboucas},
  {del Burgo}, {Graham}, {Guzman}, {Hudson}, {Matkovic}, {Merritt}, {Miller},
  {Mouhcine}, {Phillipps}, {Sharples}, {Smith}, {Tully}, \& {Verdoes
  Kleijn}}]{peng11}
{Peng}, E.~W., {Ferguson}, H.~C., {Goudfrooij}, P., {et~al.} 2011, ArXiv
  e-prints

\bibitem[{{Pryor} \& {Meylan}(1993)}]{pryor93}
{Pryor}, C. \& {Meylan}, G. 1993, in Astronomical Society of the Pacific
  Conference Series, Vol.~50, Structure and Dynamics of Globular Clusters, ed.
  {S.~G.~Djorgovski \& G.~Meylan}, 357--+

\bibitem[{{Richtler} {et~al.}(1992){Richtler}, {Grebel}, {Domgoergen},
  {Hilker}, \& {Kissler}}]{richtler92}
{Richtler}, T., {Grebel}, E.~K., {Domgoergen}, H., {Hilker}, M., \& {Kissler},
  M. 1992, \aap, 264, 25

\bibitem[{{Romanowsky} {et~al.}(2009){Romanowsky}, {Strader}, {Spitler},
  {Johnson}, {Brodie}, {Forbes}, \& {Ponman}}]{romanowsky09}
{Romanowsky}, A.~J., {Strader}, J., {Spitler}, L.~R., {et~al.} 2009, \aj, 137,
  4956

\bibitem[{{S{\'a}nchez-Bl{\'a}zquez} {et~al.}(2006){S{\'a}nchez-Bl{\'a}zquez},
  {Peletier}, {Jim{\'e}nez-Vicente}, {Cardiel}, {Cenarro},
  {Falc{\'o}n-Barroso}, {Gorgas}, {Selam}, \& {Vazdekis}}]{sanchez06}
{S{\'a}nchez-Bl{\'a}zquez}, P., {Peletier}, R.~F., {Jim{\'e}nez-Vicente}, J.,
  {et~al.} 2006, \mnras, 371, 703

\bibitem[{{Sand} {et~al.}(2008){Sand}, {Treu}, {Ellis}, {Smith}, \&
  {Kneib}}]{sand08}
{Sand}, D.~J., {Treu}, T., {Ellis}, R.~S., {Smith}, G.~P., \& {Kneib}, J. 2008,
  \apj, 674, 711

\bibitem[{{Sand} {et~al.}(2004){Sand}, {Treu}, {Smith}, \& {Ellis}}]{sand04}
{Sand}, D.~J., {Treu}, T., {Smith}, G.~P., \& {Ellis}, R.~S. 2004, \apj, 604,
  88

\bibitem[{{Schlegel} {et~al.}(1998){Schlegel}, {Finkbeiner}, \&
  {Davis}}]{schlegel98}
{Schlegel}, D.~J., {Finkbeiner}, D.~P., \& {Davis}, M. 1998, \apj, 500, 525

\bibitem[{{Schuberth} {et~al.}(2008){Schuberth}, {Richtler}, {Bassino}, \&
  {Hilker}}]{schuberth08}
{Schuberth}, Y., {Richtler}, T., {Bassino}, L., \& {Hilker}, M. 2008, \aap,
  477, L9

\bibitem[{{Schuberth} {et~al.}(2010){Schuberth}, {Richtler}, {Hilker},
  {Dirsch}, {Bassino}, {Romanowsky}, \& {Infante}}]{schuberth10}
{Schuberth}, Y., {Richtler}, T., {Hilker}, M., {et~al.} 2010, \aap, 513, A52+

\bibitem[{{Tortora} {et~al.}(2010){Tortora}, {Napolitano}, {Romanowsky}, \&
  {Jetzer}}]{tortora10}
{Tortora}, C., {Napolitano}, N.~R., {Romanowsky}, A.~J., \& {Jetzer}, P. 2010,
  \apjl, 721, L1

\bibitem[{{Ventimiglia} {et~al.}(2008){Ventimiglia}, {Arnaboldi}, \&
  {Gerhard}}]{venti08}
{Ventimiglia}, G., {Arnaboldi}, M., \& {Gerhard}, O. 2008, Astronomische
  Nachrichten, 329, 1057

\bibitem[{{Ventimiglia} {et~al.}(2011){Ventimiglia}, {Arnaboldi}, \&
  {Gerhard}}]{venti11}
{Ventimiglia}, G., {Arnaboldi}, M., \& {Gerhard}, O. 2011, ArXiv e-prints

\bibitem[{{Ventimiglia} {et~al.}(2010){Ventimiglia}, {Gerhard}, {Arnaboldi}, \&
  {Coccato}}]{venti10}
{Ventimiglia}, G., {Gerhard}, O., {Arnaboldi}, M., \& {Coccato}, L. 2010, \aap,
  520, L9+

\bibitem[{{Wehner} {et~al.}(2008){Wehner}, {Harris}, {Whitmore}, {Rothberg}, \&
  {Woodley}}]{wehner08}
{Wehner}, E.~M.~H., {Harris}, W.~E., {Whitmore}, B.~C., {Rothberg}, B., \&
  {Woodley}, K.~A. 2008, \apj, 681, 1233

\bibitem[{{Yamasaki} {et~al.}(2002){Yamasaki}, {Ohashi}, \&
  {Furusho}}]{yamasaki02}
{Yamasaki}, N.~Y., {Ohashi}, T., \& {Furusho}, T. 2002, \apj, 578, 833

\end{thebibliography}

\end{document}